\documentclass{article}
\usepackage{amsmath,amssymb,amsthm}
\usepackage{pstricks,pst-node,pst-coil}

\usepackage{graphics}

\newcommand{\bareta}{{\bar{\eta}}}

\newcommand{\inter}{{\mathrm{int}}}

\newcommand{\action}{{\cal{A}^\inter}}

\newcommand{\tr}{{\mathrm{tr}}}
\newcommand{\ee}{{\mathrm{e}}}
\newcommand{\dd}{{\mathrm{d}}}

\newcommand{\grad}{{\mathrm{deg}}}

\newcommand{\bfr}{{\mathbf{r}}}

\newcommand{\calA}{{\cal A}}

\newcommand{\Wzero}{{W^0}}

\renewcommand{\i}[1]{{}_{\scriptscriptstyle(#1)}}

\newcommand{\Deltau}{{\underline{\Delta}}}

\newcommand{\iu}[1]{{}_{\scriptscriptstyle(\underline #1)}}

\newcommand{\barpsi}{\psi^\dagger}

\newtheorem{prop}{Proposition}[section]

\begin{document}
%
%
\title{\textbf{A differential identity for Green functions}}
\author{Christian Brouder\\
Institut de Min\'eralogie et de Physique des Milieux Condens\'es, 
\\ CNRS UMR7590, Universit\'es Paris 6 et 7, IPGP, 4 place Jussieu,
\\
F-75252 Paris Cedex 05, France. \\
}%
\maketitle
\begin{abstract}
If $P$ is a differential operator with constant coefficients,
an identity is derived to calculate the action of
$\ee^P$ on the product of two functions. In many-body theory,
$P$ describes the interaction Hamiltonian and the
identity yields a hierarchy of Green functions.
The identity is first derived for scalar fields and
the standard hierarchy is recovered. Then the case of
fermions is considered and the identity is used to
calculate the generating function for the Green functions
of an electron system in a time-dependent external potential.
\end{abstract}



\section{Introduction}
In spite of immense progress in the calculation of the properties
of materials, some of them are still beyond our computational
power. Systems containing both localized and delocalized
electrons often belong to these hard problems.
A particularly conspicuous example is the 
optical spectrum of molecules and solids containing
transition metals. Brute force calculations
involving a large number of configurations do not give 
satisfactory results for the colour of such familiar materials
as blood \cite{Nakatsuji} or grass \cite{Hasegawa}.
In the solid state, non-perturbative Green function methods 
using the Bethe-Salpeter equation proved accurate for the colour
of semiconductors \cite{Albrecht,Benedict,Rohlfing}
 but not for transition metal compounds.
On the other hand, the very simple ligand field formalism can be
used to calculate the position of the spectral lines \cite{TanabeSugano},
 but not their intensities.
The success of non-perturbative Green function methods for
semiconductors and of ligand field theory for the spectral line
position makes it desirable to unify the Green function
and the ligand field theories. The present paper describes
a tool developed for this unification.

The main assumption of the Green function theory is that the
eigenstate of the interacting system can be obtained 
as the adiabatic evolution of the ground state of the
non-interacting system and that the latter can be written
as a Slater determinant. A similar assumption is present
in the Kohn-Sham approach to the density functional theory.
On the contrary, the effectiveness of the ligand field
approach comes from the fact that \emph{several} initial
states are taken into account.  The ground state of the
interacting system is then obtained by diagonalising an effective
Hamiltonian representing the effect of the electron-electron interaction
on these initial states \cite{Griffith}.
The starting point of our unification is therefore to 
use non-perturbative Green function methods to set up 
an effective Hamiltonian for a small number of
initial states. The standard Green function method is
recovered when the number of initial states is one,
the ligand field method is obtained as the first 
approximation of a hierarchy of equations for the
Green functions \cite{BrouderICDIM}.

To develop this theory, we used a mathematical tool
initiated by Schwinger \cite{SchwingerJMP} and
that was taken up by a few authors \cite{Fujita,Hall,Kukharenko}.
This approach is known as the theory
of Green functions for open 
shells \cite{NomsGFOS}
and its development was rather slow because its combinatorial
complexity is much larger than in the usual many-body
theory. 
From a graphical point of view, the complexity comes
from the fact that the Feynman diagrams must be drawn
not only with the usual two-leg Feynman propagator but also with
$2k$-leg ``propagators'' where $k$ goes from 1
to the number of initial states. These many-leg propagators
describe the correlation between initial states.
It turns out that quantum group methods are well suited
to tame this complexity and help deriving an identity
that yields a hierarchy of Green functions for open shells
\cite{BrouderEuroLett}.
More precisely, if $P$ is a differential operator
with constant coefficients,
the identity gives a closed formula for the
action of $\ee^P$ on the product of two functions 
$fg$.

For motivation, we show how this identity is used in 
many-body theory.
The Green functions $G(x_1,\dots,x_n)$ of an interacting
system can be obtained from a generating function $Z(j)$
by functional derivatives with respect to the external source 
$j(x)$ (\cite{Itzykson} p.212)
\begin{eqnarray*}
G(x_1,\dots,x_n) &=& \frac{i^n}{Z} \frac{\delta^n Z}{\delta j(x_1)\dots\delta
j(x_n)}.
\end{eqnarray*}
Therefore, in principle, $Z(j)$ provides a complete information
on the system. The generating function $Z(j)$ is usually given
by an equation of the type
$Z(j)=\ee^{P} \ee^{\Wzero}$, where $P$ is a polynomial
in $\delta/\delta j(x)$ (the functional derivative with respect to the
external source) and $\Wzero$ is a polynomial in $j$.
In standard (i.e. closed shell) Green function theory, $\Wzero$ is bilinear
in $j$:
\begin{eqnarray*}
\Wzero &=& \frac{i}{2} \int \dd x \dd y
  j(x) G^0(x,y) j(y),
\end{eqnarray*}
where $G^0(x,y)$ is the Green function of the non-interacting 
system. For the Green functions of open shells, the
function $\Wzero(j)$ is a polynomial in $j$ of degree larger
than 2. This polynomial describes the initial state of the
non-interacting system \cite{BrouderPRA}.
When we take the functional derivative of $Z(j)$
with respect to $j(x)$ and we denote 
$\delta Z(j)/\delta j(x)$ by $Z'(j)$, we find
\begin{eqnarray*}
Z'(j) &=& \frac{\delta}{\delta j(x)}
  \big(\ee^{P} \ee^{\Wzero}\big)
=\ee^P\Big(\frac{\delta \ee^{\Wzero}}{\delta j(x)}\Big)
=\ee^P\big(W'_0 \ee^{\Wzero}\big),
\end{eqnarray*}
where $W'_0=\delta \Wzero(j)/\delta j(x)$.
The expression $\ee^P$ is an infinite sum
of differential operators $P^n/n!$ acting on the product 
$W'_0 \ee^{\Wzero}$. The role of the identity is to 
transform it into a \emph{finite} sum of differential operators
acting on $\ee^{P} \ee^{\Wzero}=Z(j)$. In other words, we
obtain a non-trivial expression relating $Z'(j)$ to a differential
operator $D$ acting on $Z(j)$. This is indeed a hierarchy
of Green functions, i.e. a relation between the one-point interacting
Green function $Z'(j)$ and some $n$-point interacting
Green functions generated by $D Z(j)$.
Such a hierarchy is important because it provides 
non-perturbative equations that can be used to derive
non-perturbative approximations \cite{BrouderEuroLett}.
In ref. \cite{BrouderEuroLett}, the identity was given
without proof and no detail was given to explain how the
hierarchy follows from the identity.
The purpose of the present paper is to fill this gap by
proving the main identity and by deriving in detail some
of its applications.

The plan of the paper is as follows. After this introduction
we show that the action of a differential operator $P$ on
a product of two functions $fg$ can be described by
a mathematical concept called a \emph{coproduct}.
Then we derive our main result which provides a way
to calculate $\ee^P(fg)$. 
As a simple application of the basic identity, we derive the
hierarchy of Green functions for standard quantum field theory
of the scalar field.
We also describe the changes that are required when
we consider fermions instead of bosons. As an application,
we derive in detail the generating function $Z$ in the
presence of an external time-dependent potential.
We conclude with some possible extensions of the basic
identity to non-commutative algebras.

\section{The coproduct of a differential operator}
In this section we introduce the coproduct of a differential
operator with constant coefficients. 
We start with the case of a one-dimensional problem and
generalize it to the $d$-dimensional case.
But first, we define the tensor product of functions
and of differential operators, that will be crucial for
the definition of the coproduct.

\subsection{Tensor product}
If $f$ and $g$ are functions of a variable
$x$, the tensor product of $f$ and $g$, denoted by 
$f\otimes g$, is a function of two variable 
defined by $(f\otimes g)(x,y)=f(x)g(y)$.
From this definition, it follows that the
tensor product is linear: for any complex number
$\lambda$ and for any functions
$f$, $g$ and $h$ we have
$f\otimes (g + h)=f\otimes g + f \otimes h$,
$(f+g)\otimes h=f\otimes h + g \otimes h$
and $\lambda (f\otimes g)=(\lambda f)\otimes g=f\otimes (\lambda g)$.
Moreover, we can multiply the two tensor products
$f\otimes g$ and $f'\otimes g'$ by the rule
$(f\otimes g)(f'\otimes g')=(ff')\otimes (gg')$,
where $(ff')(x)=f(x)f'(x)$ and
$(gg')(y)=g(y)g'(y)$.

If the variable $x$ runs over a $d$-dimensional
space, we call differential operator
a finite sum 
\begin{eqnarray*}
P=\sum_{n_1,\dots,n_d} a_{n_1\dots n_d} 
  \frac{\partial^{n_1}}{\partial x_1^{n_1}}\dots
  \frac{\partial^{n_d}}{\partial x_d^{n_d}},
\end{eqnarray*}
 where $a_{n_1\dots n_d}$ are complex numbers.
If $P$ and $Q$ are differential operators acting on
functions of $x$, the 
tensor product of $P$ and $Q$, denoted by $P\otimes Q$
is the operator acting on the tensor product of two functions 
$f\otimes g$ by 
\begin{eqnarray*}
(P\otimes Q) (f\otimes g)(x,y) &=&
\big((Pf)\otimes (Qg)\big)(x,y)=
(Pf)(x) (Qg)(y).
\end{eqnarray*}
With this definition, it can be checked that the tensor
product of differential operators is linear and that
we can define the multiplication of tensor products of operators
by $(P\otimes Q)(R\otimes S)=(PR)\otimes (QS)$.
The tensor product is a familiar object in quantum physics
to define an operator acting on a many-body state.
For instance, if $L$ is the angular momentum acting on a
single particle, the angular momentum acting on two-particle
wavefunctions is defined as
$L\otimes 1+1\otimes L$.

\subsection{The one-dimensional case}
If $\partial=\dd/\dd x$, the action of $\partial^n$
on the product of two functions of $x$ is given
by the Leibniz identity
\begin{eqnarray*}
\partial^n(fg) &=&
\sum_{k=0}^n \binom{n}{k} 
  (\partial^k f)(\partial^{n-k} g).
\end{eqnarray*}
In this definition, the functions $f$ and $g$ are
not really important and look more like dummy arguments.
The coproduct is a way to calculate $\partial^n(fg)$ without
mentioning $f$ and $g$. The coproduct is denoted by
$\Delta$ and defined by
\begin{eqnarray*}
\Delta \partial^n &=& \sum_{k=0}^n \binom{n}{k} 
\partial^k \otimes \partial^{n-k}.
\end{eqnarray*}
For example
\begin{eqnarray*}
\Delta 1 &=& 1 \otimes 1,\\
\Delta \partial &=& \partial \otimes 1 + 1 \otimes\partial,\\
\Delta \partial^2 &=& \partial^2 \otimes 1 + 2 \partial \otimes \partial
+ 1 \otimes\partial^2.
\end{eqnarray*}
More precisely, the relation between the coproduct and
the derivative of $fg$ is given by 
\begin{eqnarray*}
\partial^n(fg)(x) &=& (\Delta \partial^n)(f\otimes g)(x,x).
\end{eqnarray*}
At this stage, we need to define more accurately
the algebraic structure we are using.
Let $\calA$ be the algebra of differential operators
with constant coefficients
(i.e. the algebra of polynomials in the
variable $\partial$). Thus $P\in\calA$ if and only
if there is a finite number of complex numbers
$a_n$ such that $P=\sum_{n\ge0} a_n \partial^n$.
The product in $\calA$ is the usual product of polynomials:
if $P=\sum_{n\ge0} a_n \partial^n$ and
$Q=\sum_{m\ge0} b_m \partial^m$ then
$PQ=\sum_{n,m\ge0} a_nb_m \partial^{n+m}$.
Thus, $\calA$ is a unitary commutative algebra.
The coproduct appears when the differential operators
act on a product of two functions.
It is a linear map $\Delta : \calA \rightarrow \calA\otimes\calA$
whose action on an element $P\in\calA$ can be written
$\Delta P = \sum_i R_i \otimes S_i$, so that
$P(fg)=\sum_i (R_i f) (S_i g)$.
For example, for $P=\partial^2$ the coproduct is a sum
of three terms described by
$R_1=\partial^2$, $S_1=1$, $R_2=2\partial $, $S_2=\partial$
and $R_3=1$, $S_3=\partial^2$ (note that, by linearity of the
tensor product, we can also choose
$R_2=\partial $, $S_2=2\partial$).
In general, we can use          
\begin{eqnarray*}
P\big(fg\big) &=&
\sum_{n\ge0} a_n \sum_{k=0}^n \binom{n}{k} 
  (\partial^k f)(\partial^{n-k} g),
\end{eqnarray*}
to get
\begin{eqnarray}
\Delta P &=&
\sum_{n\ge0} a_n \sum_{k=0}^n \binom{n}{k} 
  \partial^k \otimes \partial^{n-k}.
\label{DeltaP}
\end{eqnarray}
In other words, the coproduct is just a way
to denote which differential operators act on
$f$ and on $g$ in $P(fg)$. The introduction of such
a concept may look rather pedantic at this stage,
but it will allow us to obtain
very general results, valid for partial derivatives
and even for functional derivatives.
Note that the $k$-th derivative of $\partial^n$
with respect to $\partial$ is $n! \partial^{n-k}/(n-k)!$
for $n\ge k$ and $0$ for $n<k$,
so that 
\begin{eqnarray}
\Delta P &=&
\sum_{k\ge 0}\frac{\partial^k}{k!} 
  \otimes P^{(k)},
\label{DeltaP2}
\end{eqnarray}
where $P^{(k)}$ is the $k$-th derivative of 
$P$ with respect to $\partial$.
The concept of a coproduct is unfamiliar, but it is very
useful in quantum theory. For instance, it was needed to
calculate matrix elements of many-body operators and 
density correlations \cite{BrouderPRA}.
In this paper we shall see that it is also quite
useful to derive non-perturbative equations in
Green function theory.
We hope that the example of the coproduct of
a differential operator makes this concept
understandable.
For convenience, we shall replace the 
notation
$\Delta P = \sum_i R_i \otimes S_i$ by
Sweedler's notation
$\Delta P = \sum P\i1 \otimes P\i2$,
which is more common.

\subsection{The $d$-dimensional case}
In the $d$-dimensional case, we put
$\partial_i=\partial/\partial x_i$ and we define the coproduct
of $D=\prod_{i=1}^d \partial_i^{n_i}$ as
\begin{eqnarray*}
\Delta D &=& 
\sum_{k_1=0}^{n_1} \dots
\sum_{k_d=0}^{n_d} 
\binom{n_1}{k_1}\dots \binom{n_d}{k_d}
\partial_1^{k_1}\dots \partial_d^{k_d}
\otimes 
\partial_1^{n_1-k_1}\dots \partial_d^{n_d-k_d}.
\end{eqnarray*}

We can even define the coproduct of an infinite dimensional
differential operator. If 
$\bfr_1,\dots,\bfr_p$ are $p$ points, 
the coproduct of the functional differential operator
\begin{eqnarray*}
D &=& 
\frac{\delta^{n_1+\dots+n_p}}
{\delta j(\bfr_1)^{n_1}\dots\delta j(\bfr_p)^{n_p}}
=
\frac{\delta^{n_1}}
{\delta j(\bfr_1)^{n_1}} \dots
\frac{\delta^{n_p}} {\delta j(\bfr_p)^{n_p}}
\end{eqnarray*}
is defined by
\begin{eqnarray*}
\Delta D &=& 
\sum_{k_1=0}^{n_1} \dots
\sum_{k_p=0}^{n_p} 
\binom{n_1}{k_1}\dots \binom{n_p}{k_p}
\frac{\delta^{k_1+\dots+k_p}}
{\delta j(\bfr_1)^{k_1}\dots\delta j(\bfr_p)^{k_p}}
\otimes 
\frac{\delta^{n_1-k_1+\dots+n_p-k_p}}
{\delta j(\bfr_1)^{n_1-k_1}\dots\delta j(\bfr_p)^{n_p-k_p}}.
\end{eqnarray*}

\subsection{Algebra morphism}
Up to now, the coproduct is just a linear map from
$\calA$ to $\calA\otimes\calA$. The coproduct becomes
a powerful tool if it has a property called 
\emph{algebra morphism}. The coproduct is an algebra
morphism if $\Delta 1=1\otimes 1$ and, 
for any $P$ and $Q$ in $\calA$,
$\Delta (PQ)=(\Delta P)(\Delta Q)$ or, more
explicitly
\begin{eqnarray*}
\Delta (PQ) &=& \sum (PQ)\i1 \otimes (PQ)\i2
= \sum (P\i1 Q\i1) \otimes (P\i2 Q\i2)
\\&=&
 \big(\sum P\i1 \otimes P\i2 \big)
 \big(\sum Q\i1 \otimes Q\i2 \big).
\end{eqnarray*}

To show that the coproduct of the algebra of
differential operators is an algebra morphism,
we consider for notational convenience the
one-dimensional case.
If we take $P=\partial^n$ and $Q=\partial^m$,
then $PQ=\partial^{n+m}$ and
\begin{eqnarray*}
\Delta (PQ) &=& 
  \sum_{i=0}^{m+n} \binom{n+m}{i} \partial^i \otimes \partial^{n+m-i}
=
  \sum_{k=0}^{n}\sum_{l=0}^m \binom{n}{k} \binom{m}{l}
  \partial^{k+l} \otimes \partial^{n+m-k-l}
\\&=&
  \Big( \sum_{k=0}^{n}\binom{n}{k}
  \partial^{k} \otimes \partial^{n-k}\Big)
  \Big( \sum_{l=0}^m \binom{m}{l}
  \partial^{l} \otimes \partial^{m-l}\Big)
 =(\Delta P)(\Delta Q),
\end{eqnarray*}
where we used the Vandermonde convolution formula
$\sum_{k+l=i}
\binom{n}{k}\binom{m}{l}=\binom{n+m}{i}$.
The $d$-dimensional case is proved similarly.

\subsection{Reduced coproduct}
For any element $P$ of $\calA$, the reduced coproduct
of $P$ is defined by
\begin{eqnarray*}
\Deltau P &=& \Delta P - P\otimes 1- 1\otimes P.
\end{eqnarray*}
For example:
$\Deltau \partial =0$, 
$\Deltau \partial^2 =2\partial\otimes\partial$, 
\begin{eqnarray*}
\Deltau \partial^n &=& \sum_{k=1}^{n-1} \binom{n}{k} 
\partial^k \otimes \partial^{n-k},
\end{eqnarray*}
for $n>1$.
We denote the reduced coproduct of $P$ by
$\Deltau P = \sum P\iu1\otimes P\iu2$.
If the polynomial $P$ is such that $P(0)=0$, 
its coproduct $\Delta P$ contains the terms
$P\otimes 1$ and $1\otimes P$.
The purpose of the reduced coproduct is to eliminate
these terms. As a result of this operation,
the degree of $P\iu1$ and $P\iu2$ is always greater than 0
and smaller than the degree of $P$.

\section{The main identity}
In this section, we prove the identity from which
we can derive, among other things, the hierarchy of Green functions.
Although this identity provides a powerful method to
resum infinities of  Feynman diagrams of the perturbation
theory, its proof is very simple.
\begin{prop}
\label{mainthm}
If $\calA$ is a commutative algebra equipped with a 
coproduct $\Delta$ that is an algebra morphism and if
$P\in\calA$ then
\begin{eqnarray}
\Delta (\ee^{P}) &=&  
\ee^{\Deltau P}\big(\ee^{P} \otimes \ee^{P}\big),
\label{DeltaeP}
\end{eqnarray}
where
$\Deltau P=\Delta P - P\otimes 1 - 1 \otimes P$ and
\begin{eqnarray*}
\ee^{\Deltau P} &=&  
1\otimes 1 + \sum_{n=1}^\infty \frac{1}{n!} (\Deltau P)^n
=
1\otimes 1 + \sum_{n=1}^\infty \frac{1}{n!} \Big(\sum P\iu1\otimes P\iu2\Big)^n
\end{eqnarray*}
\end{prop}
\begin{proof}
By linearity of the coproduct, we have
\begin{eqnarray*}
\Delta (\ee^{P}) &=& 
\Delta(\sum_{n=0}^\infty \frac{1}{n!} P^n) 
=
\sum_{n=0}^\infty \frac{1}{n!} \Delta(P^n) 
=
\sum_{n=0}^\infty \frac{1}{n!} (\Delta P)^n
=\ee^{\Delta P},
\end{eqnarray*}
where we used the fact that
$\Delta$ is an algebra morphism
to write $\Delta(P^n)=(\Delta P)^n$.
In $\ee^{\Delta P}$, the 
operator $M=\Delta P$ is considered as a linear operator
from $\calA\otimes\calA$ to $\calA\otimes\calA$ defined by
$M(Q\otimes R)=\sum (P\i1 Q)\otimes (P\i2 R)$.
We decompose this operator $M$
as $M=\Deltau P + P\otimes 1 + 1\otimes P$ and we use the fact
that the algebra $\calA$ is commutative to write
\begin{eqnarray*}
\ee^M &=& \ee^{\Deltau P + P\otimes 1 + 1\otimes P}
=\ee^{\Deltau P}\ee^{P\otimes 1}\ee^{1\otimes P}.
\end{eqnarray*}
The proof is completed by noting that
\begin{eqnarray*}
\ee^{P\otimes 1} &=& 
\sum_{n=0}^\infty \frac{1}{n!} (P\otimes 1)^n 
=
\sum_{n=0}^\infty \frac{1}{n!} P^n\otimes 1 =\ee^P \otimes 1.
\end{eqnarray*}
Similarly, $\ee^{1\otimes P}=1\otimes \ee^P$, so that
$\ee^{P\otimes 1}\ee^{1\otimes P}=(\ee^P \otimes 1)(1\otimes \ee^P)
=\ee^P \otimes \ee^P$.
\end{proof}
Note that $\Deltau$ is not an algebra morphism
so that $(\Deltau P)^n\not= \Deltau (P^n)$. Thus,
we will use a special notation to write the terms of
$(\Deltau P)^n$. Namely,
\begin{eqnarray*}
(\Deltau P)^n &=& \sum P\i{1'}^n \otimes  P\i{2'}^n.
\end{eqnarray*}
With this notation, we can write the action of 
$\ee^P$ on the product of two functions as
\begin{eqnarray}
\ee^P(fg) &=& (\ee^P f)(\ee^P g) + \sum_{n=1}^\infty
  \frac{1}{n!} \sum (P\i{1'}^n \ee^P f) (P\i{2'}^n \ee^P g).
\label{ePfg}
\end{eqnarray}

\section{Applications}
We consider now some applications of the identity
(\ref{ePfg}) of increasing complexity.
The first application is a simple proof that
$\ee^{a\partial} f(x)=f(x+a)$, the second is 
a hierarchy of Green functions for closed shells.

\subsection{The shift operator}
For this very simple application, assume that you do not
know Taylor's theorem and that you want to evaluate
$\ee^{a\partial} f(x)$ for an entire function $f$.
To prove that $\ee^{a\partial} f(x)=f(x+a)$ it is enough
to show that $\ee^{a\partial} x^n=(x+a)^n$.
The proof will be recursive.
For $n=1$, $\partial^0 x=x$,
$\partial^1 x=1$ and $\partial^k x=0$ for $k>1$.
Thus, $\ee^{a\partial} x=(1+a\partial)x=x+a$.
Now take $n>1$ and $P=a\partial$.
We have $\Delta P= P \otimes 1 + 1 \otimes P$,
so that $\Deltau P=0$ and the identity (\ref{ePfg}) gives us
$\ee^P(fg)=(\ee^P f)(\ee^P g)$. Therefore,
taking $f(x)=x^{n-1}$ and $g(x)=x$ we have
$\ee^P(x^n)=(\ee^P x^{n-1})(\ee^P x)$. The recursion hypothesis
tells us that $\ee^P x^{k}=(x+a)^{k}$ for all $k<n$, thus
$\ee^P(x^n)=(x+a)^{n-1}(x+a)=(x+a)^n$.

\subsection{The standard hierarchy}
In the quantum theory of the scalar field,
the system is described by the action
$A(\varphi)=A_0(\varphi)-V(\varphi)$, with
\begin{eqnarray*}
A_0(\varphi) &=& 
  \frac{1}{2} \int\dd x \partial_\mu\varphi(x) \partial^\mu\varphi(x)
  - m^2 \varphi^2(x),
\end{eqnarray*}
and $V(\varphi)=\int\dd x F(\varphi(x))$, where 
$F(\varphi(x))$ is a polynomial such that $F(0)=F'(0)=0$
(usually $\varphi^3(x)/3!$ or $\varphi^4(x)/4!$).
The generating function
of the Green functions of this theory can be obtained from the
following formula (see \cite{Itzykson} p. 445)
\begin{eqnarray*}
Z(j) &=& \ee^{-iV(-i\delta_j)} \ee^{\frac{i}{2} \int \dd x \dd y
  j(x) G^0(x,y) j(y)},
\end{eqnarray*}
where $\delta_{j}$ is a short notation for the functional
derivative $\delta/\delta j(x)$, and where $G^0(x,y)$
is the Green function of the free scalar field.

A hierarchy of Green functions is an exact relation between interacting
Green functions. To obtain it, we first take the functional
derivative of $Z(j)$ with respect to $j(x)$:
\begin{eqnarray*}
\frac{\delta Z(j)}{\delta j(x)} &=& \ee^{-iV(-i\delta_j)} 
  \frac{\delta }{\delta j(x)}
     \ee^{\frac{i}{2} \int \dd z \dd z' j(z) G^0(z,z') j(z')}
\\&=&
 i\ee^{-iV(-i\delta_j)} 
\Big(\int \dd y G^0(x,y) j(y)
     \ee^{\frac{i}{2} \int \dd z \dd z' j(z) G^0(z,z') j(z')}
  \Big).
\end{eqnarray*}
Now we use equation (\ref{ePfg}) with
$f=\int \dd y G^0(x,y) j(y)$,
$g=\ee^{\frac{i}{2} \int \dd z \dd z' j(z) G^0(z,z') j(z')}$
and $P=-iV(-i\delta_j)$.
We first calculate $(\ee^P f)=\ee^{-iV(-i\delta_j)} \int \dd y G^0(x,y) j(y)$.
From the fact that $F(0)=F'(0)=0$, we see that no terms of
$F$ has a degree smaller than 2. Therefore,
$V(-i\delta_j)  \int \dd y G^0(x,y) j(y)=0$ because
$\int \dd y G^0(x,y) j(y)$ is of degree 1 in $j(x)$.
This gives us 
\begin{eqnarray*}
\ee^{-iV(-i\delta_j)} \int \dd y G^0(x,y) j(y) &=&
\int \dd y G^0(x,y) j(y).
\end{eqnarray*}
Now we calculate the reduced coproduct $\Deltau V$.
If $F(\varphi(x))$ is a polynomial in 
$\varphi(x)$, $F(-i\delta_j)$ is a polynomial 
in $\delta_j$. Thus
\begin{eqnarray*}
\Delta V &=& \int \dd y \sum_{k\ge0} \frac{1}{k!}
  \delta_{j(y)}^k \otimes \frac{\partial^k F}{\partial \delta_{j(y)}^k}
=
\int \dd y \sum_{k\ge0} \frac{(-i)^k}{k!}
  \delta_{j(y)}^k \otimes F^{(k)}(-i\delta_{j(y)}).
\end{eqnarray*}
Therefore,
\begin{eqnarray*}
\Deltau V &=& 
\int \dd y \sum_{k\ge1} \frac{(-i)^k}{k!}
  \delta_{j(y)}^k \otimes F^{(k)}(-i\delta_{j(y)}) - V\otimes 1.
\end{eqnarray*}
The term $n=1$ of the main identity (\ref{ePfg}) becomes
\begin{eqnarray*}
\int \dd y \sum_{k\ge1} \frac{(-i)^k}{k!}
  \big(\delta_{j(y)}^k f\big)
  \big(F^{(k)}(-i\delta_{j(y)}) \ee^P g\big)
 - (Vf)(\ee^P g).
\end{eqnarray*}
The fact that $f=\int \dd y G^0(x,y) j(y)$ is of degree 1
in $j$ implies that only the term $k=1$ may give a
non-zero contribution.  Its value is
\begin{eqnarray*}
\int \dd y (-i)
  \big(\delta_{j(y)} f\big)
  \big(F'(-i\delta_{j(y)}) \ee^P g\big)
&=&
-i \int \dd y G^0(x,y) F'(-i\delta_{j(y)}) Z(j).
\end{eqnarray*}
Moreover, all terms coming from 
$(\Deltau V)^n$ with $n>1$ in the main identity 
(\ref{ePfg}) give no contribution
because they are of degree at least $n$ in $\delta_j$.
This gives us the simple result
\begin{eqnarray}
\frac{\delta Z(j)}{\delta j(x)} &=& 
i \int \dd y G^0(x,y) j(y) Z(j)
-i \int \dd y G^0(x,y) F'(-i\delta_{j(y)}) Z(j).
\label{hierscal}
\end{eqnarray}
If we multiply this equation by $(\Box+m^2)$ we get
the standard hierarchy (see \cite{Itzykson} p. 447)
\begin{eqnarray}
(\Box+m^2)\frac{\delta Z(j)}{\delta j(x)} &=& 
i j(x) Z(j)
-i F'(-i\delta_{j(x)}) Z(j),
\label{hierscaldif}
\end{eqnarray}
which is usually obtained by path-integral methods.
However, note that equation (\ref{hierscaldif})
does not imply equation (\ref{hierscal})
but
\begin{eqnarray*}
\frac{\delta Z(j)}{\delta j(x)} &=& 
i \int \dd y G^0(x,y) j(y) Z(j)
-i \int \dd y G^0(x,y) F'(-i\delta_{j(y)}) Z(j)
+\phi(x),
\end{eqnarray*}
where $\phi(x)$ is a solution
of the scalar wave equation $(\Box+m^2)\phi(x)=0$.
Thus, equation (\ref{hierscal}) is a result
stronger than (\ref{hierscaldif}).

\section{The fermionic case}
For applications to molecular or solid-state physics,
we need to modify the equations to treat the case of
fermions.
The main difference with the previous case is
in the definition of the tensor product.

\subsection{Tensor product}
The basic variables of the fermionic theories are
the field operators $\psi(x)$ and $\barpsi(x)$
and the fermionic sources $\eta(x)$ and $\bareta(x)$.
These variables are assumed to anticommute:
for example $\eta(x)\psi(y)=-\psi(y)\eta(x)$.

The functions $f$ will be polynomials in these
basic variables.
If $f$ is a monomial in the basic variables,
we denote by $\grad(f)$ the degree of $f$.
For example, if $f$ is a basic variable, $\grad(f)=1$. For
$f=\barpsi(\bfr)\psi(\bfr)$, $\grad(f)=2$; for
$f=\int \bareta(x)\psi(x) \dd x$, $\grad(f)=2$.
The parity of a function $f$, denoted by $|f|$,
is 0 if $\grad(f)$ is even and is 1 if $\grad(f)$ is odd.
Moreover, a function $f$ is said to be even if
$|f|=0$ and odd if $|f|=1$.
From the anticommutation of the basic variables, one can 
derive the following commutation relation of two functions
$f$ and $g$: $gf=(-1)^{|f||g|} fg$.
Therefore, an even function commutes with
all functions and two odd functions anticommute.

The differential operators are products of functional
derivatives with respect to $\eta$ or 
$\bareta$. The basic relations are
\begin{eqnarray*}
\frac{\delta}{\delta\eta(x)}\eta(y) &=& \delta(x-y),
\quad
\frac{\delta}{\delta\eta(x)}\bareta(y) = 0,
\nonumber\\
\frac{\delta}{\delta\bareta(x)}\bareta(y) &=& \delta(x-y),
\quad
\frac{\delta}{\delta\bareta(x)}\eta(y) = 0.
\end{eqnarray*}
If $f$ and $g$ are functions in the basic variable
with a definite parity, then the functional derivative
of the product $fg$ is given by the modified 
Leibniz formula
\begin{eqnarray}
\frac{\delta}{\delta\eta(x)} (fg) &=& \frac{\delta f}
     {\delta\eta(x)}g
+(-1)^{|f|} f \frac{\delta g}{\delta\eta(x)},
\label{Leibniz}
\end{eqnarray}
and the same equation for a functional derivative with
respect to $\bareta(x)$.
Equation (\ref{Leibniz}) is known as Leibniz' rule.

The sources $\eta$ and $\bareta$ anticommute, so
the functional derivatives anticommute:
\begin{eqnarray*}
\frac{\delta^2}{\delta\eta(x)\delta\eta(y)} &=& 
-\frac{\delta^2}{\delta\eta(y)\delta\eta(x)}.
\end{eqnarray*}

The degree of a monomial 
$P$ in $\delta/\delta \eta(x)$ 
and $\delta/\delta \bareta(x)$ is denoted by 
$\grad(P)$ and the parity of $P$, denoted by 
$|P|$ is 0 if $\grad(P)$ is even
and 1 if $\grad(P)$ is odd.
The action of a differential operator on a product of
two functions is given by
\begin{eqnarray*}
P(fg) &=& \sum (-1)^{|f||P\i2|} (P\i1 f) (P\i2 g),
\end{eqnarray*}
where the coproduct of terms of degree 1 is
\begin{eqnarray*}
\Delta \frac{\delta}{\delta \eta(x)} &=&
\frac{\delta}{\delta \eta(x)} \otimes 1
+ 1 \otimes \frac{\delta}{\delta \eta(x)},\\
\Delta \frac{\delta}{\delta \bareta(x)} &=&
\frac{\delta}{\delta \bareta(x)} \otimes 1
+ 1 \otimes \frac{\delta}{\delta \bareta(x)},
\end{eqnarray*}
and the coproduct of a product of
two monomials $P$ and $Q$ is calculated from the
coproduct of $P$ and $Q$ by using the fact that
the coproduct is an algebra morphism:
\begin{eqnarray}
\Delta (PQ) &=& (\Delta P)(\Delta Q)
=\sum (-1)^{|P\i2||Q\i1|} (P\i1 Q\i1)\otimes (P\i2 Q\i2),
\label{almorfer}
\end{eqnarray}
where we have used the fact that the multiplication of 
$P\otimes Q$ by $R\otimes S$ is
$(-1)^{|Q||R|} (PR)\otimes (QS)$.

\subsection{The main identity}
With the conventions described in the last paragraph,
the main identity (\ref{DeltaeP}) is valid without
change for fermionic sources if $P$ is even (i.e. if $|P|=0$).
The action on a product of two functions is now
\begin{eqnarray}
\ee^P (fg) &=& (\ee^P f)(\ee^P g) +
  \sum_{n=1}^\infty \frac{1}{n!} 
\sum (-1)^{|f||P\i{2'}^{n}|} \big(P\i{1'}^{n} 
  (\ee^P f)\big)
\big(P\i{2'}^{n} (\ee^P g)\big).
\label{ePABfermion}
\end{eqnarray}

\section{Application of the fermionic formula}
The main application of the fermionic formula
was the derivation of the hierarchy of Green
functions for open shells \cite{BrouderEuroLett}.
It would be too long to give a detailed derivation of 
this hierarchy. So we consider the calculation
of the generating function $Z$ of the Green functions
for a system in a time-dependent external potential.
This generating function is useful 
in time-dependent density functional theory.
We first describe some properties of fermionic
sources, then we give the step by step derivation
of the dependence of $Z$ on $\eta$ and $\bareta$,
using the main identity. Finally, we use this identity
again to obtain the exact expression of $Z$.

\subsection{Fermion sources}
The S-matrix of a fermionic 
system with external sources $\eta$ and $\bareta$
and action $\action$ is
\begin{eqnarray*}
S(\bareta,\eta) &=& 
T \exp\big(-i\action +i\int \bareta(x)\psi(x) \dd x
  +i \int \barpsi(x)\eta(x) \dd x \big).
\end{eqnarray*}
For a nonrelativistic fermion, $\psi(x)$ and
$\barpsi(x)$ have a spin index. Thus,
the sources are also two-component vectors
and 
\begin{eqnarray*}
\bareta(x)\psi(x) &=& \sum_{s=1}^2
  \bareta_s(x)\psi_s(x),
\quad\quad
\barpsi(x)\eta(x) = \sum_{s=1}^2
  \barpsi_s(x)\eta_s(x). 
\end{eqnarray*}

To see how functional derivatives act with
respect to the time-ordering operator, we
first notice that the sources
can be taken out of the time-ordering operator.
For example, if $x^0>y^0$
\begin{eqnarray*}
T(\bareta(x)\psi(x)\barpsi(y)\eta(y))
&=& \bareta(x)\psi(x)\barpsi(y)\eta(y) =
\bareta(x)\eta(y) \psi(x)\barpsi(y)
\\&=& \bareta(x)\eta(y) T(\psi(x)\barpsi(y)),
\end{eqnarray*}
if $x^0<y^0$
\begin{eqnarray*}
T(\bareta(x)\psi(x)\barpsi(y)\eta(y))
&=& \barpsi(y)\eta(y) \bareta(x)\psi(x)=
-\bareta(x)\eta(y) \barpsi(y)\psi(x)
\\&=& \bareta(x)\eta(y) T(\psi(x)\barpsi(y)).
\end{eqnarray*}
Thus, the functional derivative with respect to
$\eta(x)$ or $\bareta(x)$ commutes with the time-ordering
operator.
In particular  \cite{Itzykson}
\begin{eqnarray*}
\frac{\delta S(\bareta,\eta)}{\delta\bareta(x)}|_{\bareta=\eta=0} &=&
i T \big(\psi(x)\ee^{-i\action}\big)
,\\
\frac{\delta S(\bareta,\eta)}{\delta\eta(x)}|_{\bareta=\eta=0} &=&
-i T \big(\barpsi(x)\ee^{-i\action}\big),
\end{eqnarray*}
where the minus sign in the
last equation comes from the fact that the functional derivative
must jump over $\barpsi(x)$ to reach $\eta(x)$
in the definition of $S(\bareta,\eta)$.

A standard result of the functional derivative approach
\cite{Itzykson,Redmond} is that
the interacting S-matrix $S(\bareta,\eta)$ can
be obtained from the non-interacting S-matrix
\begin{eqnarray*}
S^0(\bareta,\eta) &=&
T \exp\big(i \int \bareta(x)\psi(x) + \barpsi(x)\eta(x)\dd x \big)
\end{eqnarray*}
by the equation
\begin{eqnarray*}
S(\bareta,\eta) &=& 
\exp\big(-i\int_{-\infty}^\infty 
     H^\inter(\frac{-i\delta}{\delta\bareta(x)},
               \frac{i\delta}{\delta\eta(x)}) \dd t\big)
               S^0(\bareta,\eta),
\end{eqnarray*}
where $H^\inter(\psi(\bfr,t),\barpsi(\bfr,t))$
is the interaction Hamiltonian.

If $|\Phi_0\rangle$ is a non-degenerate eigenstate of $H_0$,
the Green functions of the interacting
system can be obtained from the generating function
\begin{eqnarray*}
Z(\bareta,\eta) &=& \langle \Phi_0| S(\bareta,\eta) |\Phi_0\rangle
=
\exp\big(-i\int_{-\infty}^\infty 
     H^\inter(\frac{-i\delta}{\delta\bareta(x)},
               \frac{i\delta}{\delta\eta(x)}) \dd t\big)
               Z^0(\bareta,\eta),
\end{eqnarray*}
where $Z^0=\langle \Phi_0| S^0 |\Phi_0\rangle$.
If  $|\Phi_0\rangle$
can be written as a Slater determinant,
$|\Phi_0\rangle = b^\dagger_{i_N}\dots b^\dagger_{i_1} |0\rangle$,
where $b^\dagger_n$ is the creation operator of the
one-electron state $u_n(x)$ of the free Hamiltonian, 
we have \cite{BrouderPRA}
\begin{eqnarray*}
Z^0(\bareta,\eta) &=&
\langle \Phi_0 |T \exp\big(i \int \bareta(x)\psi(x) 
   + \barpsi(x)\eta(x)\dd x \big)
|\Phi_0\rangle =
 \ee^{-i\int \bareta(x) G^0(x,y)\eta(y)\dd x\dd y},
\end{eqnarray*}
where
\begin{eqnarray*}
G^0(x,y) &=&
-i\langle \Phi_0 | T\big(\psi(x)\barpsi(y)\big)|\Phi_0\rangle
= -i\langle 0 | T\big(\psi(x)\barpsi(y)\big)|0\rangle
  +i\sum_{k=1}^N u_{i_k}(x) u^*_{i_k}(y).
\end{eqnarray*}

As a useful example, we consider the interaction with an external time-dependent
potential $v(x)$ so that
$\action = \int \dd x \barpsi(x) v(x) \psi(x)$.
The generating function becomes
$Z=\ee^P \ee^{\Wzero}$, where 
\begin{eqnarray*}
P &=& -i \int \dd x v(x) \frac{\delta^2}{\delta \eta(x)\delta
\bareta(x)},\\
\Wzero &=& 
 -i\int \bareta(x) G^0(x,y)\eta(y)\dd x\dd y.
\end{eqnarray*}
Our purpose is now to calculate $Z$.

\subsection{Dependence of $Z$ on $\eta$ and $\bareta$}
For later convenience, we consider 
$Z(\lambda)=\ee^P \ee^{\lambda \Wzero}$.
We first calculate the derivative of $Z(\lambda)$ with respect to
$\bareta(x)$.
\begin{eqnarray*}
\frac{\delta Z(\lambda)}{\delta \bareta(x)}
&=& \ee^P \Big(\frac{\delta \ee^{\lambda \Wzero}}{\delta \bareta(x)}\Big)
= \lambda \ee^P \Big(\frac{\delta \Wzero}{\delta \bareta(x)} 
   \ee^{\lambda \Wzero}\Big)
= -i \lambda \ee^P \Big(
\int G^0(x,y)\eta(y)\dd y \ee^{\lambda \Wzero}\Big).
\end{eqnarray*}
We can apply identity (\ref{ePABfermion}) with
$f=\int G^0(x,y)\eta(y)\dd y$, so that
$|f|=1$ and 
$g= \ee^{\lambda \Wzero}$. We first note that
$P f=0$ because $f$ does not contain any factor $\bareta$.
Thus $\ee^P f=f$ and since $\ee^P g=Z(\lambda)$,
equation (\ref{ePABfermion}) becomes
\begin{eqnarray*}
\frac{\delta Z(\lambda)}{\delta \bareta(x)}
 &=& -i\lambda f Z(\lambda) -i\lambda
  \sum_{n=1}^\infty \frac{1}{n!} 
\sum (-1)^{|P\i{2'}^{n}|} \big(P\i{1'}^{n} 
  f\big)
\big(P\i{2'}^{n} Z(\lambda)\big).
\end{eqnarray*}
To proceed, we must calculate the reduced coproduct of $P$.
We first have the coproduct
\begin{eqnarray*}
\Delta P &=&  P\otimes 1 + 1 \otimes P
-i \int \dd z v(z) 
   \frac{\delta}{\delta \eta(z)}\otimes
   \frac{\delta}{\delta \bareta(z)}
+i \int \dd z v(z) 
   \frac{\delta}{\delta \bareta(z)}\otimes
   \frac{\delta}{\delta \eta(z)},
\end{eqnarray*}
so that
\begin{eqnarray}
\Deltau P &=&
-i \int \dd z v(z) 
   \frac{\delta}{\delta \eta(z)}\otimes
   \frac{\delta}{\delta \bareta(z)}
+i \int \dd z v(z) 
   \frac{\delta}{\delta \bareta(z)}\otimes
   \frac{\delta}{\delta \eta(z)}.
\label{DeltauP}
\end{eqnarray}
The term $f$ is of degree 1 in $\eta$ and 0 in $\bareta$.
Thus, the factor $P\i{1'}^{n} f$ is non zero only
if $n=1$ and $P\i{1'}=\delta/\delta\eta(z)$. We find
\begin{eqnarray*}
\frac{\delta Z(\lambda)}{\delta \bareta(x)}
 &=& -i\lambda f Z(\lambda) -i\lambda
(-i) \int \dd z v(z) (-1) 
   \frac{\delta f}{\delta \eta(z)}
   \frac{\delta Z(\lambda)}{\delta \bareta(z)}
\\&=&
  -i\lambda f Z(\lambda) +\lambda
\int \dd y v(y) G^0(x,y)
   \frac{\delta Z(\lambda)}{\delta \bareta(y)}.
\end{eqnarray*}
If we write $Z(\lambda)=\ee^{W(\lambda)}$ we get
\begin{eqnarray*}
\frac{\delta W(\lambda)}{\delta \bareta(x)}
 &=& 
  -i\lambda \int \dd y G^0(x,y)\eta(y)  +\lambda
\int \dd y G^0(x,y) v(y)
   \frac{\delta W(\lambda)}{\delta \bareta(y)},
\end{eqnarray*}
with the solution 
\begin{eqnarray}
\frac{\delta W(\lambda)}{\delta \bareta(x)}
 &=& 
  -i\lambda \int \dd y G(x,y)\eta(y),
\label{soldWbareta}
\end{eqnarray}
where $G(x,y)$ satisfies the Dyson equation
\begin{eqnarray}
G(x,y) &=& G^0(x,y)
 + \lambda \int \dd z G^0(x,z)v(z) G(z,y).
\label{DysonG}
\end{eqnarray}
In other words, $G(x,y)$ is the Green function of the
Schr\"odinger equation obtained by adding 
$\lambda v$ to the free Hamiltonian.
From equation (\ref{soldWbareta}), we see that
the function $W(\lambda)$ can be written
$W(\lambda)
= 
  -i\lambda \int \dd x \dd y \bareta(x) G(x,y)\eta(y)
  + g(\eta,\lambda)$.
The same calculation for $\delta Z(\lambda)/\delta\eta(y)$
shows that
$\delta W(\lambda)/\delta \eta(y) = 
  i\lambda \int \dd x \bareta(x) G(x,y)$.
Thus, $g$ does not depend on $\eta$ and
\begin{eqnarray}
W(\lambda)
 &=& 
  -i\lambda \int \dd x \dd y \bareta(x) G(x,y)\eta(y)
  + g(\lambda),
\label{Wlambda1}
\end{eqnarray}
where $g(\lambda)$ is independent of $\bareta$ and $\eta$.

\subsection{Calculation of $g(\lambda)$}
To calculate $g(\lambda)$, we take
the derivative of $Z(\lambda)$ with respect to
$\lambda$.
\begin{eqnarray*}
\frac{\partial Z(\lambda)}{\partial \lambda}
&=& \ee^P \Big(\frac{\partial \ee^{\lambda \Wzero}}{\partial \lambda}\Big)
= \ee^P \Big(\Wzero \ee^{\lambda \Wzero}\Big).
\end{eqnarray*}
We can apply identity (\ref{ePABfermion}) with
$f=\Wzero$, so that $|f|=0$ and 
$g= \ee^{\lambda \Wzero}$. We now have
\begin{eqnarray*}
P \Wzero &=& -i 
\int \dd x v(x) \frac{\delta^2 \Wzero}{\delta \eta(x)\delta
\bareta(x)}
=
-\int v(x) G^0(x,x)\dd x
=-\tr(vG^0).
\end{eqnarray*}
Thus $\ee^P f=f-\tr(vG^0)$ and since $\ee^P g=Z(\lambda)$,
equation (\ref{ePABfermion}) becomes
\begin{eqnarray*}
\frac{\partial Z(\lambda)}{\partial \lambda}
 &=& \Wzero Z(\lambda) -\tr(vG^0)Z(\lambda)+
  \sum_{n=1}^\infty \frac{1}{n!} 
\sum \big(P\i{1'}^{n} 
   \Wzero\big)
\big(P\i{2'}^{n} Z(\lambda)\big).
\end{eqnarray*}
To calculate the term $n=1$, we use the reduced coproduct of $P$
given in equation (\ref{DeltauP})
and we obtain
\begin{eqnarray*}
\sum \big(P\i{1'} 
   \Wzero\big)
\big(P\i{2'} Z(\lambda)\big) &=&
-i \int \dd z v(z) 
   \frac{\delta \Wzero}{\delta \eta(z)}
   \frac{\delta Z(\lambda)}{\delta \bareta(z)}
+i \int \dd z v(z) 
   \frac{\delta \Wzero}{\delta \bareta(z)}
   \frac{\delta Z(\lambda)}{\delta \eta(z)}
\\&&\hspace*{-17mm}=
\int \dd x \dd y \bareta(x) G^0(x,y) v(y) 
   \frac{\delta Z(\lambda)}{\delta \bareta(y)}
+\int \dd x \dd y v(x) G^0(x,y) \eta(y)
   \frac{\delta Z(\lambda)}{\delta \eta(x)}.
\end{eqnarray*}
For the term $n=2$ we first evaluate, using algebra morphism
(\ref{almorfer})
\begin{eqnarray*}
(\Deltau P)^2 &=&
\Big(-i \int \dd x v(x) 
   \frac{\delta}{\delta \eta(x)}\otimes
   \frac{\delta}{\delta \bareta(x)}
+i \int \dd x v(x) 
   \frac{\delta}{\delta \bareta(x)}\otimes
   \frac{\delta}{\delta \eta(x)}\Big)
\\&&\hspace*{3mm}
\Big(-i \int \dd y v(y) 
   \frac{\delta}{\delta \eta(y)}\otimes
   \frac{\delta}{\delta \bareta(y)}
+i \int \dd y v(y) 
   \frac{\delta}{\delta \bareta(y)}\otimes
   \frac{\delta}{\delta \eta(y)}\Big)
\\&=&
  \int\dd x\dd y v(x) v(y)
   \frac{\delta^2}{\delta \eta(x)\delta \eta(y)}\otimes
   \frac{\delta^2}{\delta \bareta(x)\delta\bareta (y)}
\\&&
 -
  \int\dd x\dd y v(x) v(y)
   \frac{\delta^2}{\delta \eta(x)\delta \bareta(y)}\otimes
   \frac{\delta^2}{\delta \bareta(x)\delta\eta (y)}
\\&&
 -
  \int\dd x\dd y v(x) v(y)
   \frac{\delta^2}{\delta \bareta(x)\delta \eta(y)}\otimes
   \frac{\delta^2}{\delta \eta(x)\delta\bareta (y)}
\\&&
 +
  \int\dd x\dd y v(x) v(y)
   \frac{\delta^2}{\delta \bareta(x)\delta \bareta(y)}\otimes
   \frac{\delta^2}{\delta \eta(x)\delta\eta (y)}.
\end{eqnarray*}
$\Wzero$ contains one $\eta$ and one $\bareta$. Thus, the only
terms of $(\Deltau P)^2$ that give non-zero contributions
are the second and the third, because the left hand side of 
the tensor product contains functional
derivatives with respect to $\eta$ and $\bareta$.
If we interchange the variables $x$ and $y$, we see
that these two terms are identical, thus
the term $n=2$ becomes
\begin{eqnarray*}
  \frac{1}{2} 
\sum \big(P\i{1'}^{2} 
   \Wzero\big)
\big(P\i{2'}^{2} Z(\lambda)\big)
&=& - \int\dd x\dd y v(x) v(y)
   \frac{\delta^2 \Wzero}{\delta \eta(y)\delta \bareta(x)}
   \frac{\delta^2 Z(\lambda)}{\delta \bareta(y)\delta\eta (x)}
\\&=&
i  \int\dd x\dd y v(x) v(y) G^0(x,y)
   \frac{\delta^2 Z(\lambda)}{\delta \bareta(y)\delta\eta (x)}.
\end{eqnarray*}
There is no term for $n>2$ because $\Wzero$ is of degree 2.

This gives us the equation for $Z(\lambda)$
\begin{eqnarray*}
\frac{\partial Z(\lambda)}{\partial \lambda}
 &=& \Wzero Z(\lambda) -\tr(vG^0)Z(\lambda)+
\int \dd x \dd y \bareta(x) G^0(x,y) v(y) 
   \frac{\delta Z(\lambda)}{\delta \bareta(y)}
\\&&
+\int \dd x \dd y v(x) G^0(x,y) \eta(y)
   \frac{\delta Z(\lambda)}{\delta \eta(x)}
+ i  \int\dd x\dd y v(x) v(y) G^0(x,y)
   \frac{\delta^2 Z(\lambda)}{\delta \bareta(y)\delta\eta (x)}.
\end{eqnarray*}
If we write $Z(\lambda)=\ee^{W(\lambda)}$ we obtain the
equation for $W(\lambda)$
\begin{eqnarray*}
\frac{\partial W(\lambda)}{\partial \lambda}
 &=& 
 -i\int \bareta(x) G^0(x,y)\eta(y)\dd x\dd y
 -\tr(vG^0)+
\int \dd x \dd y \bareta(x) G^0(x,y) v(y) 
   \frac{\delta W(\lambda)}{\delta \bareta(y)}
\\&&
+\int \dd x \dd y v(x) G^0(x,y) \eta(y)
   \frac{\delta W(\lambda)}{\delta \eta(x)}
+ i  \int\dd x\dd y v(x) v(y) G^0(x,y)
   \frac{\delta W(\lambda)}{\delta \bareta(y)}
   \frac{\delta W(\lambda)}{\delta\eta (x)}
\\&&
+ i  \int\dd x\dd y v(x) v(y) G^0(x,y)
   \frac{\delta^2 W(\lambda)}{\delta \bareta(y)\delta\eta (x)}.
\end{eqnarray*}
The general form of $W(\lambda)$ is given by equation
(\ref{Wlambda1}). If we introduce it into the last equation
we obtain
\begin{eqnarray*}
-i \bareta G\eta
-i \lambda \bareta \frac{\partial G}{\partial \lambda} \eta
+g'(\lambda)
 &=& 
 -i \bareta G^0\eta
 -\tr(vG^0)-i\lambda
\bareta G^0 v G\eta
-i\lambda \bareta G v G^0 \eta
\\&&
- i \lambda^2  \bareta G v G^0 v G \eta
- \lambda\tr(v G^0  v G).
\end{eqnarray*}
This gives us two independent equations:
\begin{eqnarray*}
G + \lambda \frac{\partial G}{\partial \lambda}
 &=& 
  G^0 +\lambda G^0 v G
+\lambda  G v G^0
+ \lambda^2  G v G^0 v G,\\
g'(\lambda) &=&
 -\tr(vG^0) - \lambda \tr(v G^0  v G).
\end{eqnarray*}
Equation (\ref{DysonG}) enables us to simplify this into
\begin{eqnarray*}
G + \lambda \frac{\partial G}{\partial \lambda} &=& 
  G +\lambda  G v G,\\
g'(\lambda) &=& -\tr(vG).
\end{eqnarray*}
The first equation is an identity because
the Dyson equation  (\ref{DysonG}) yields
\begin{eqnarray*}
\frac{\partial G}{\partial \lambda} &=& 
G^0 v G + \lambda G^0 v \frac{\partial G}{\partial \lambda},
\end{eqnarray*}
so that 
\begin{eqnarray*}
\frac{\partial G}{\partial \lambda} &=& G v G.
\end{eqnarray*}
The last equation to solve is simply
$g'(\lambda) = -\tr(vG)=-i\int v(x)\rho(x;\lambda)\dd x$, where
$\rho(x;\lambda)=-i G(x,x)$ is the charge density in the presence
of the potential $\lambda v$.
The solution of this equation would be
$g(\lambda)-g(0)=\int_0^\lambda g'(\mu)\dd\mu$.
If $\lambda=0$, $Z(0)=\ee^P 1=1$. Thus, $W(0)=0$.
Moreover, for $\lambda=0$, $G=G^0$ and
$W=g(0)$. Thus, $g(0)=0$.
Now we prove that 
\begin{eqnarray}
g(\lambda) &=& -\tr\big(\log(1+\lambda vG)\big).
\label{glambda}
\end{eqnarray}
By definition, the right hand side of this equation is
\begin{eqnarray*}
-\tr\big(\log(1+\lambda vG)\big) 
&=& \sum_{n=1}^\infty \frac{(-\lambda)^n}{n} \tr\big((vG)^n\big).
\end{eqnarray*}
If we take the derivative with respect to $\lambda$ we find
\begin{eqnarray*}
-\frac{\partial \tr\big(\log(1+\lambda vG)\big) }{\partial \lambda}
&=& -\sum_{n=1}^\infty (-\lambda)^{n-1}\tr\big((vG)^n\big)+
\sum_{n=1}^\infty (-\lambda)^n 
  \tr\Big(v\frac{\partial G}{\partial \lambda}(vG)^{n-1}\Big)
\\&=&
 -\sum_{n=0}^\infty (-\lambda)^{n}\tr\big((vG)^{n+1}\big)+
\sum_{n=1}^\infty (-\lambda)^n 
  \tr\big((vG)^{n+1}\big)=-\tr(vG).
\end{eqnarray*}
Therefore, $g(\lambda)$ and $-\tr\big(\log(1+\lambda vG)\big)$
satisfy the same first order differential equation.
Moreover, the boundary conditions are the same because
$-\tr\big(\log(1)\big)=0=g(0)$.

\subsection{Alternative expressions}
We can give alternative expressions for $g(\lambda)$.
Expanding $G=G^0+\lambda G^0vG$ we obtain
$G=G^0\sum_{n=0}^\infty (\lambda vG^0)^n$
so that
\begin{eqnarray*}
g'(\lambda) &=& - \tr(vG)
=-\sum_{n=0}^\infty \lambda^n \tr\big((vG^0)^{n+1}\big).
\end{eqnarray*}
Hence,
\begin{eqnarray*}
g(\lambda) &=& \int_0^\lambda g'(\mu)\dd\mu 
=-\sum_{n=0}^\infty \frac{\lambda^{n+1}}{n+1} \tr\big((vG^0)^{n+1}\big)
= \tr\log(1-\lambda v G^0).
\end{eqnarray*}
A similar result was obtained by Sham \cite{Sham85} in the
framework of the density functional theory.
The relation with equation (\ref{glambda}) is obtained from
$G=G^0+\lambda G v G^0$, so that
$vG^0=vG(1-\lambda vG^0)$.
Therefore, formally,
\begin{eqnarray*}
\tr\log(1-\lambda v G^0) &=&
\tr\log(vG^0)-\tr\log(vG)
=\tr\log(vG^0)-\tr\log(vG^0(1+\lambda v G))
\\&=&
-\tr\log(1+\lambda v G).
\end{eqnarray*}
We can manipulate these operators as if they were scalars
because $vG$ is a function of $vG^0$, so that every term
is a series in the single variable $vG^0$.

To give a last form of $g(\lambda)$, we define the Euler operator
$A=\int \dd x v(x)\frac{\delta}{\delta v(x)}$, that counts
the number of times $v$ is present in an expression. 
In other words, it is easy to check that
$A \tr\big((vG^0)^n\big)= n \tr\big((vG^0)^n\big)$.
More generally,
$A^k \tr\big((vG^0)^n\big)= n^k \tr\big((vG^0)^n\big)$.
Thus, we can write the somewhat formal identity
\begin{eqnarray*}
g(\lambda) &=& 
-\sum_{n=0}^\infty \frac{\lambda^{n+1}}{n+1} \tr\big((vG^0)^{n+1}\big)
= - A^{-1} \sum_{n=0}^\infty \lambda^{n+1}\tr\big((vG^0)^{n+1}\big)
=- \lambda  A^{-1} \tr(vG).
\end{eqnarray*}

\section{Conclusion}
The hierarchy of Green functions is one of the building blocks of
an extension of the Bethe-Salpeter equation to open shells.
The other main ingredient is an expression of the Green functions
in terms of 2-particle-irreducible generating functions.
This expression is well known for closed shells
\cite{Pismak4} but not for open shells.
In the latter case, the Dyson equation relating the two-point
Green function and the self-energy (i.e. one-particle irreducible
Green function) is not valid. This is a bad news because the
Dyson equation is required to use the Legendre transformation
relating the connected Green function and the one- or two-particle
irreducible ones.
By a diagrammatic tour de force, Hall was able to obtain the equation 
corresponding to the Dyson equation for open shells \cite{Hall}. 
It looks very complex but, by doubling the size of the Green functions,
it is possible to give it a form similar to the traditional
Dyson equation. This gives some hope that the Bethe-Salpeter
equation for open shells can be obtained,
opening the way to an effective unification of the
ligand field and Green function methods.

In some applications, the operator $P$ in $\ee^P$
belongs to a noncommutative algebra \cite{Blasiak}.
This is the case for example when $P$ is a differential operator
with non-constant coefficients.
It is possible to generalize our main identity to noncommutative
algebras.  The main problem is to 
transform $\ee^{\Delta P}$ into an operator $\ee^A$
acting on $1\otimes \ee^P$. In other words, we have to solve
the equation
$\ee^A=\ee^{\Delta P} \ee^{-1\otimes P}$. This can
be done by using the Baker-Campbell-Hausdorff formula
\cite{Reutenauer}.

\section{Acknowledgements}
I am very grateful to Emanuela Petracci for her introduction
to the Hopf algebra of derivations. I thank Alessandra Frabetti
and Kurusch Ebrahimi-Fard for lively discussions on the 
extension to non-commutative algebras.


\begin{thebibliography}{10}
\bibitem{NomsGFOS}
The method is also know as Green functions with initial correlations or Green
  functions for degenerate systems. It belongs to the framework of the theory
  of non-equilibrium Green functions.
\end{thebibliography}

\end{document}